\documentclass[a4paper,12pt]{article}
\usepackage[utf8]{inputenc}
\usepackage[english]{babel}
\usepackage{amsmath, amssymb,graphicx,bm}
\usepackage{braket}
\usepackage{caption}
\usepackage{subcaption}
\usepackage{ulem}
\usepackage{multicol}
\usepackage{appendix}
\pdfoutput=1
\usepackage{jheppubm}
\newcommand{\be}{\begin{equation}}
\newcommand{\ee}{\end{equation}}
\newcommand{\bea}{\begin{eqnarray}}
\newcommand{\eea}{\end{eqnarray}}

\newcommand{\cS}{\mathcal{S}}
\newcommand{\cA}{\mathcal{A}}

\newcommand{\fb}{\mathfrak{b}}
\newcommand{\fc}{\mathfrak{c}}

\newcommand{\fg}{\mathfrak{g}}

\newcommand{\fz}{\mathfrak{z}}

\definecolor{darkraspberry}{rgb}{0.53,0.15,0.34}

\definecolor{darkblue}{rgb}{0,0,1}

\definecolor{Nikolaevcolor}{rgb}{.855,0,.855}

\definecolor{dgreen}{rgb}{0,0.6,0}

\definecolor{brown}{rgb}{0.59,0.29,0}

\definecolor{orange}{rgb}{0.89,0.31,0.05}

\numberwithin{equation}{section}

\begin{document}

\title{Holographic QCD Running Coupling for Heavy Quarks in Strong Magnetic Field
}

\author{Irina Ya. Aref'eva$^a$, Ali Hajilou$^a$, Alexander Nikolaev$^{a}$, Pavel Slepov$^a$}

\affiliation{$^a$Steklov Mathematical Institute, Russian Academy of  Sciences, \\ Gubkina str. 8, 119991, Moscow, Russia}

\emailAdd{arefeva@mi-ras.ru}
\emailAdd{hajilou@mi-ras.ru}
\emailAdd{nikolaev@mi-ras.ru}
\emailAdd{slepov@mi-ras.ru}

\abstract{We investigate the influence of a magnetic field on the running coupling constant for a heavy-quark model in a bottom-up holographic approach. To achieve this, we employ a magnetized Einstein-Maxwell-dilaton background that captures the essential features of heavy quark dynamics. Similar to the light-quark model, the running coupling $\alpha$ for heavy quarks decreases in the presence of a strong external magnetic field at fixed temperature and chemical potential. The key distinction between the light and heavy quark models lies in the locations of their respective phase transitions. However, near the 1st order phase transitions, the behavior 
  of  $\alpha$
is analogous for both cases: $\alpha$
  exhibits jumps that depend on temperature, chemical potential, and magnetic field strength.
}



\maketitle

\newpage


\section{Introduction}

The main purpose of this paper is to study the influence of the strong external magnetic field on the running coupling constant in the holographic model for heavy quarks.
The running coupling in holographic isotropic model of quantum chromodynamics (QCD) was considered in  \cite{Arefeva:2024vom}. In the paper \cite{Arefeva:2024xmg} we studied the running coupling for light-quark model in the anisotropic background, inspired by the strong external magnetic field \cite{Arefeva:2022avn,Arefeva:2020vae}, and in this research we investigate the heavy-quark system.\\

To get knowledge of interplay between phase structure of QCD and running coupling constant especially at non-zero magnetic field, that is expected to be important in the corresponding experiments, one need information on behaviour of QCD with non-zero chemical potential and external magnetic field. From theoretical point of view this means we have to have knowledge on nonperturbative effects in QCD under significant values of chemical potential, magnetic field and quark mass. As has been mentioned in numerous publications, holography proposes a non-perturbative approach to study the corresponding regimes of QCD \cite{Maldacena:1997re,Casalderrey-Solana:2011dxg, Arefeva:2014kyw,deWolf}. The lattice calculations cannot cover the finite values of the  chemical potential. In addition, perturbative methods no longer work for studying quark-gluon plasma (QGP) and the strongly coupled regime of the QCD.
Therefore, we utilize the holographic approach to investigate the effect of the magnetic field on the running coupling for heavy-quark model.\\

The QGP produced and studied in heavy ion collisions (HIC) is considered as a new phase of matter. QGP is strongly coupled system and we need holographic approach to investigate its physics. In particular, in the early stages of noncentral HIC a very strong magnetic field $eB\sim 0.3\, GeV^2$ is created \cite{Zhang:2023ppo,Skokov:2009qp,Bzdak:2011yy}. It would be interesting to study the effect of the external magnetic field on the QCD features such as phase diagram and running couplings between the quarks.
\\

To study the running coupling,  the dilaton field has a crucial role in the holographic approach, that the isotropic case was considered in \cite{Arefeva:2024vom} and the effect of strong magnetic field on the running coupling was studied in \cite{Arefeva:2024xmg}. One can apply different boundary conditions on the dilaton field that can affect the behavior of the running coupling. It is important to note that the phase structure of light and heavy quarks do not depend on the choice of different boundary conditions \cite{Arefeva:2024vom}. The effect of strong magnetic field on the running coupling considering physical boundary condition for light quark model was investigated in \cite{Arefeva:2024xmg}. In this work, it would be interesting to study the effect of strong magnetic field on the running coupling for heavy quarks model that reproduced magnetic catalysis phenomenon \cite{Arefeva:2023jjh}.\\

The energy-dependent running coupling constant in QCD remains a cornerstone of high-energy physics, with experimental determinations spanning a wide range of energies \cite{ParticleDataGroup:2024cfk} and primarily in low-density regimes (i.e., low baryon chemical potential). To elucidate the interplay between QCD’s phase structure and the running coupling under external magnetic fields, a regime of direct relevance to HIC experiments and astrophysical phenomena, a robust theoretical understanding of nonperturbative QCD dynamics at finite chemical potential, strong magnetic fields, and varying quark masses is essential. In this study, we focus on the distinctive features of heavy-quark holographic models, contrasting them with their light-quark counterparts to highlight mass-dependent effects in magnetized environments.\\

We need to emphasize that since there are some distinct differences between heavy and light quarks. One of the main differences between them is the position of the first-order phase transition in (chemical potential, temperature)-plane. In the context of holography, the warp factor in the metric distinguishes characteristics between heavy and light quarks. We utilize completely different functions as a scale factor, i.e. a polynomial for heavy quarks and logarithmic one for light quarks. These choices lead to very different outputs to describe the physics of heavy and light quarks in the context of holography. In particular, heavy-quark models show the magnetic catalysis phenomenon \cite{Arefeva:2023jjh}, and while the light-quark models show the inverse magnetic catalysis phenomenon \cite{Arefeva:2024xmg,Arefeva:2022avn}. For this reason we investigated light-quark and heavy-quark models independently. \\

\begin{figure}[h!]
\centering
\includegraphics[scale=0.19]{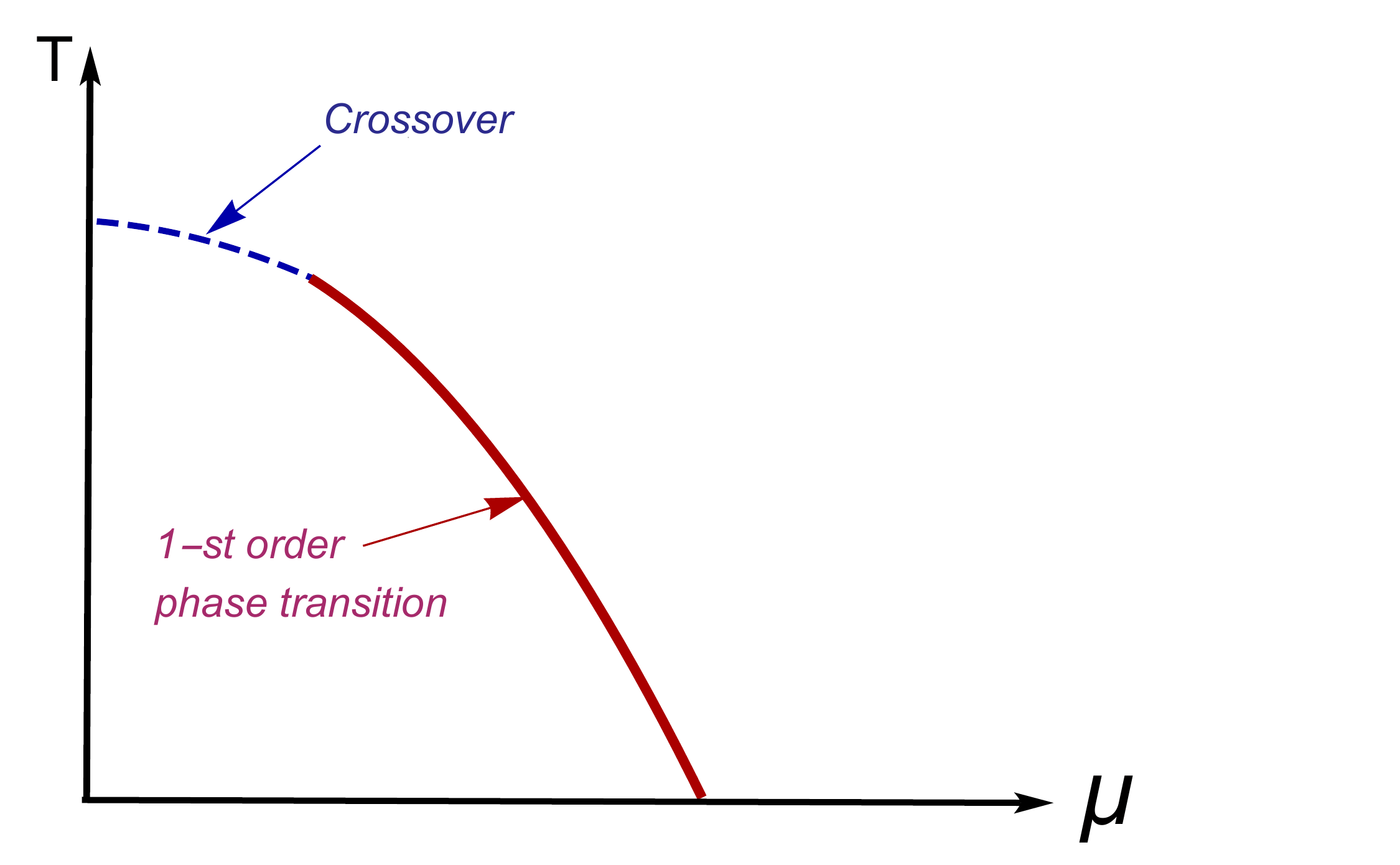} \quad
\includegraphics[scale=0.19]{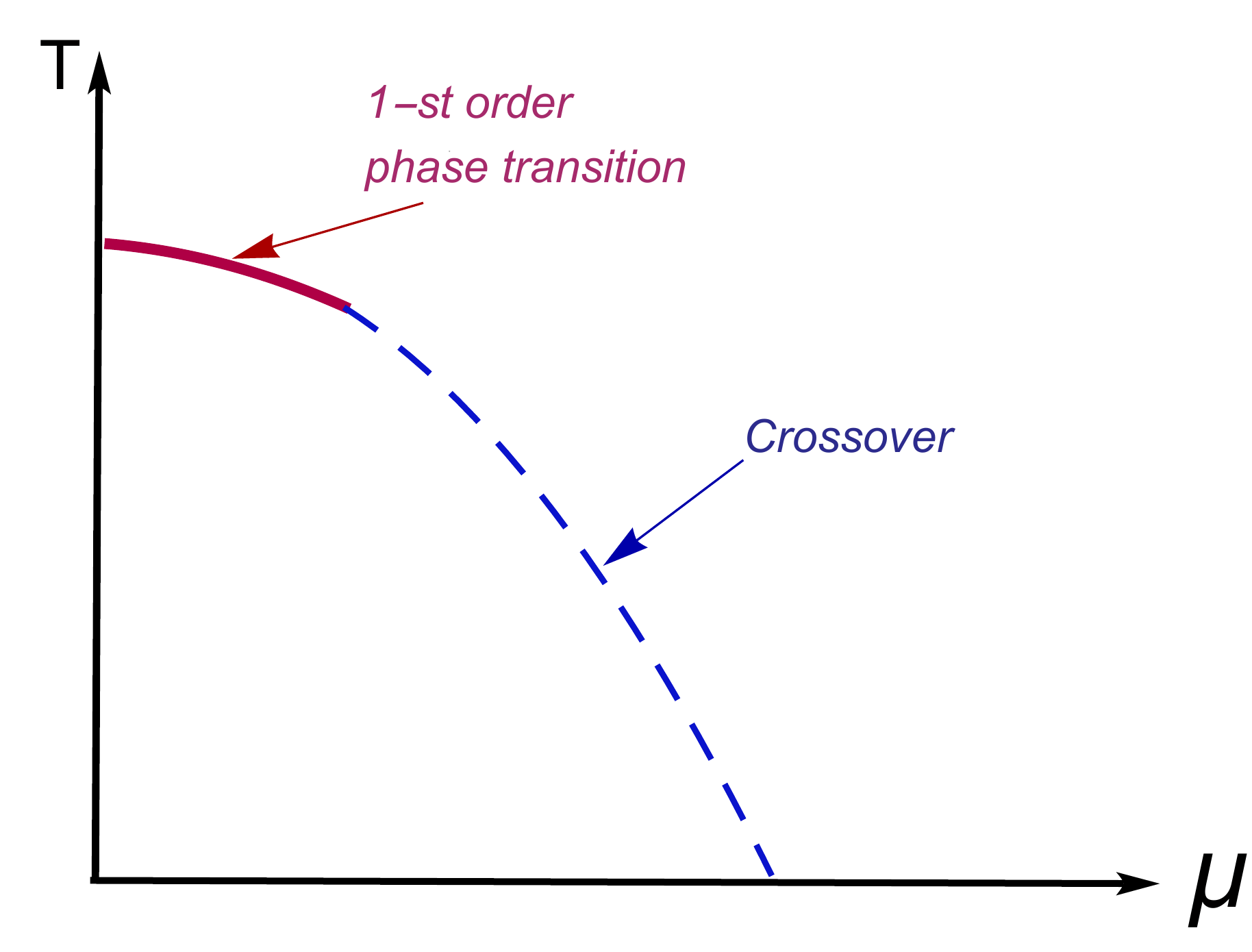} \\
  A \hspace{70 mm} B
\caption{Schematic phase diagrams in the 
$(\mu,T$)-plane for the light-quark model (A) and heavy-quark model (B).}
\label{Fig:Tmu-LQHQ}
\end{figure}

The structure of this paper is organized as follows. 
In Sect.\,\ref{sec:prelim}, we introduce the 5-dimensional fully anisotropic holographic models in the presence of a non-zero magnetic field for the heavy-quarks model.
Sect.\,\ref{sec:running}, encompasses  our results on the study of the effect of the magnetic field on the running coupling constant in the heavy-quarks model.
In Sect.\,\ref{Sect:Conc}, we summarize our numerical results obtained for heavy-quarks model by comparing them with the light-quarks model as well as  figure out perspectives to extend our results for future research.
The paper is complemented by an Appendix, which describes how the equations of motion (EOMs) were solved and which boundary conditions were applied.

\newpage

\section{ Gravitational Setup for Heavy-Quarks Model}\label{sec:prelim}

We take the 5-dimensional  Einstein-two Maxwell-dilaton system with the action in the Einstein frame, given by \cite{Arefeva:2023jjh}
\begin{gather}
  {\cS} = \int \cfrac{d^5x \, }{16\pi G_5} \sqrt{-\fg} \left[
    R - \cfrac{f_1(\varphi)}{4} \ F^{(1)^2}
        - \cfrac{f_B(\varphi)}{4} \ F^{(B)^2}
    - \cfrac{\partial_{\mu} \varphi \partial^{\mu} \varphi}{2}
    - V(\varphi) \right]\,, \label{eq:2.01}
\end{gather}
where $G_5$ is the 5-dimensional Newtonian gravitational constant, $\fg$ is the determinant of the metric tensor, $R$ is the Ricci scalar, and $\varphi = \varphi(z)$ represents the dilaton field. The functions $f_1(\varphi)$ and $f_B(\varphi)$ are the coupling functions associated with the Maxwell fields $A_{\rho}$ and $F_{\rho\sigma}^{(B)}$ ($F_{\rho\sigma}=\partial_{\rho}A_{\sigma}-\partial_{\sigma}A_{\rho}$), respectively.
The variable $\rho$ and $\sigma$ indicates the spacetime coordinates $(t, x, y_1, y_2, z)$, with $z$ being the holographic radial coordinate. Here, $V(\varphi)$ denotes the dilaton field potential.

According to the holographic dictionary, $F^{(1)}$ and $F^{(B)}$ in the gravity background correspond to the chemical potential and external magnetic field in the 
boundary field theory, respectively. Our ansatz for the non-zero components of the first and second Maxwell fields are
\begin{gather}
 A_{\rho}^{(1)} = A_t (z) \delta_\rho^t\,,
  \quad
  F_{x y_1}^{(B)} = q_B\, , \label{eq:2.02}
\end{gather}
where $A_t (z)$ is the time component of the first gauge field, and $q_B$ is a constant charge. Our ansatz for the metric is \cite{Arefeva:2023jjh}
\begin{gather}
  ds^2 = \cfrac{L^2 \, \fb(z)}{z^2} \left[
    - \, g(z) dt^2 + dx^2 
    +  \, dy_1^2
    + e^{c_B z^2}  dy_2^2
    + \cfrac{dz^2}{g(z)} \right], \label{eq:2.03} \\
  \fb(z) = e^{2{\cA}(z)}\,,  \label{eq:2.04}
\end{gather}
where $L$ is the AdS radius that we set $L=1$ in the numerical calculations and $\fb(z)$ is the warp factor. Here, $c_B$ is the magnetic field parameter that is introduced in metric to describe the strong magnetic field in noncentral HIC

It is very important to note that, the light and heavy quark models are distinguished by the scalar factor ${\cA}(z)$, defined by the warp factor \eqref{eq:2.04} in the metric \eqref{eq:2.03}. For the "heavy-quark" model a simple choice of the scale factor is ${\cA}(z) = - \fc\,z^2/4$ \cite{Andreev:2006ct,Arefeva:2018hyo}. The magnetic field has been incorporated for this model with a simple scalar factor in \cite{Bohra:2019ebj,Arefeva:2020vae}.
Here, for the heavy-quark model we consider an extended scale factor \cite{Arefeva:2023jjh}
\bea \label{scaleHQX}
\cA(z)=-\fc\,z^2/4- (p-c_B\, q_B) z^4~,
\eea
where $\fc=4 R_{gg}/3$, $R_{gg}=1.16$ GeV${}^2$, $\fc=1.547$ GeV${}^2$, and $p = 0.273$ GeV${}^4$ are parameters that can be fixed with lattice and experimental data for zero magnetic field, i.e. for the case $c_B=0$ considered in \cite{Yang:2015aia}. The main motivation for the  choice \eqref{scaleHQX} is the effect of the magnetic catalysis phenomenon \cite{Arefeva:2023jjh}.

The EOMs for this heavy quark model can be obtained as \eqref{eq:2.05}-\eqref{eq:2.10} by varying the action \eqref{eq:2.01}. All EOMs are presented in appendix \ref{app1}. In addition, to solve the EOMs it is needed to impose proper boundary conditions \eqref{eq:2.11}-\eqref{eq:2.13} that is described in appendix \ref{app1}. Note that the boundary conditions with $z_0 = 0$ has been used in \cite{Yang:2015aia}, and  \cite{Arefeva:2018hyo} discusses with $z_0 = z_h$. Here, $z_h$ denotes the size of the black hole horizon and $z_0$ is an arbitrary value of the holographic coordinate $z$.
  For our purposes, the choice of the $z_0$ as a function of $z_h$ will be discussed below.

To investigate the thermodynamics and in particular the free energy of the model, we need to evaluate the  blackening function. Solving the EOM \eqref{eq:2.07} one can obtain the blackening function $g(z)$ as

\begin{gather}
  g(z) = e^{c_B z^2} \left[ 1 - \cfrac{\Tilde{I}_1(z)}{\Tilde{I}_1(z_h)}
    + \cfrac{\mu^2 \bigl(2 R_{gg} + c_B (q_B - 1) \bigr)
      \Tilde{I}_2(z)}{L^2 \left(1 - e^{(2 R_{gg}+c_B(q_B-1))\frac{z_h^2}{2}}
      \right)^2} \left( 1 - \cfrac{\Tilde{I}_1(z)}{\Tilde{I}_1(z_h)} \,
      \cfrac{\Tilde{I}_2(z_h)}{\Tilde{I}_2(z)} \right)
  \right], \label{eq:4.42} \\
  \Tilde{I}_1(z) = \int_0^z
  e^{\left(2R_{gg}-3c_B\right)\frac{\xi^2}{2}+3 (p-c_B \, q_B) \xi^4}
  \xi^{3} \, d \xi, \qquad \ \label{eq:4.43-1} \\
  \Tilde{I}_2(z) = \int_0^z
  e^{\bigl(2R_{gg}+c_B\left(\frac{q_B}{2}-2\right)\bigr)\xi^2+3 (p-c_B
    \, q_B) \xi^4} \xi^{3} \, d \xi \,, \label{eq:4.43} 
\end{gather}
where $\mu$ is the chemical potential. Then, considering the metric in   \eqref{eq:2.03} and utilizing the extended scale
factor \eqref{scaleHQX}, the temperature and the entropy density for the heavy-quark model are given by

\begin{gather}
  \begin{split}
    T &= \cfrac{|g'|}{4 \pi} \, \Biggl|_{z=z_h} = \left|
     - \, \cfrac{e^{(2R_{gg}-c_B)\frac{z_h^2}{2}+3 (p-c_B \, q_B)
         z_h^4} \, 
      z_h^{3}}{4 \pi \, \Tilde{I}_1(z_h)} \right. \times \\
    &\left. \times \left[ 
      1 - \cfrac{\mu^2 \bigl(2 R_{gg} + c_B (q_B - 1) \bigr) 
        \left(e^{(2 R_{gg} + c_B (q_B - 1))\frac{z_h^2}{2}}\Tilde{I}_1(z_h) -
          \Tilde{I}_2(z_h) \right)}{L^2 \left(1 
          - e^{(2R_{gg}+c_B(q_B-1))\frac{z_h^2}{2}}
        \right)^2} \right] \right|, 
    \\
    s  &=\cfrac{A}{4}\Biggl|_{z=z_h}= \ \cfrac{1}{4} \left( \cfrac{L}{z_h} \right)^{3}
    e^{-(2R_{gg}-c_B)\frac{z_h^2}{2}-3 (p-c_B \, q_B) z_h^4}\, ,
  \end{split} \label{eq:4.48}
\end{gather}
where $g'$ is the derivative with respect to $z$, and $A$ is the horizon area of the black brane. Here we fixed $G_5 = 1$. Calculating the free energy
\begin{gather}
  F = - \int s \, d T = \int_{z_h}^{\infty} s \, T' \, dz\,, \label{eq:3.05}
\end{gather}
leads to obtain the phase diagrams for the heavy-quark model. Note that in \eqref{eq:3.05} the "$T'$" is "$\frac{dT}{dz}$" and we normalized the free energy considering \( z_h \to \infty \) when we have the thermal-AdS background.

\begin{figure}[h!]
  \centering
\includegraphics[scale=0.45]{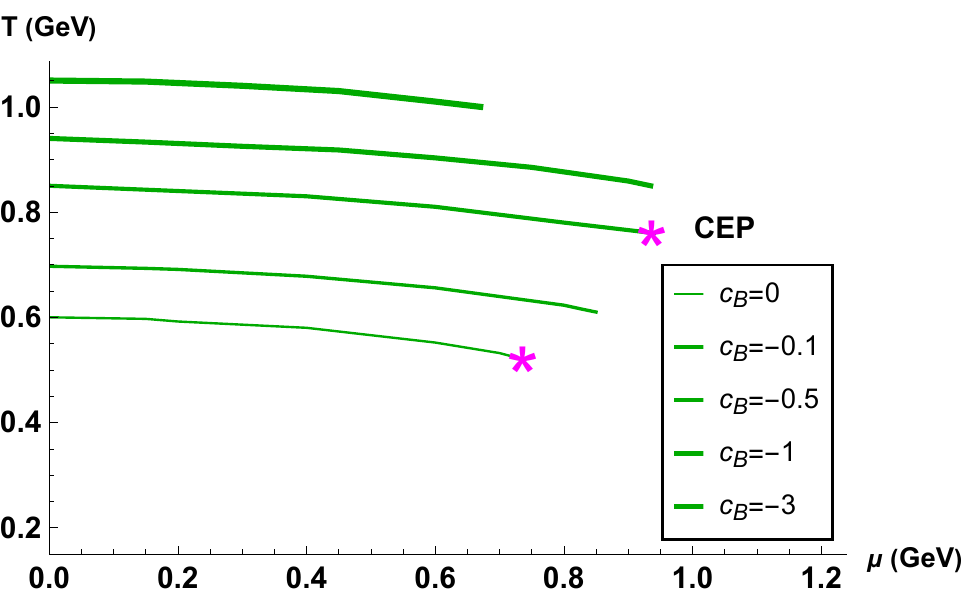} 
\includegraphics[scale=0.48]{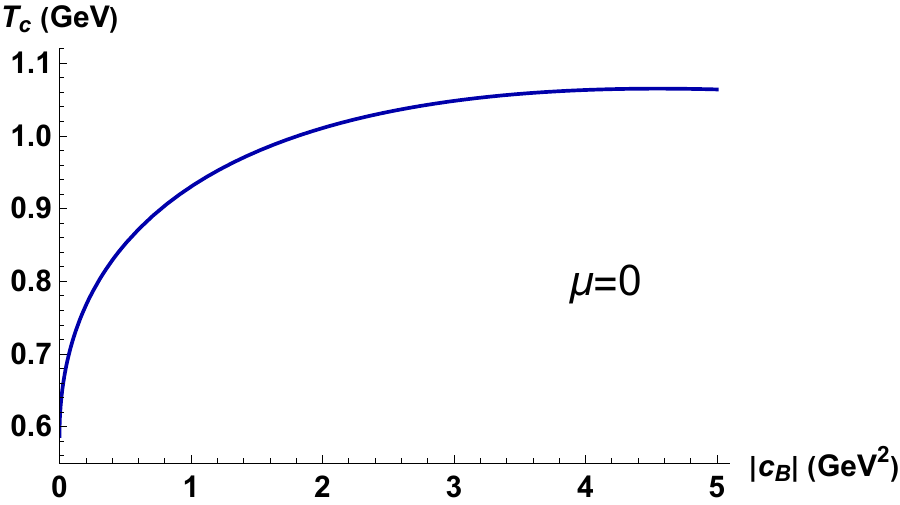}\\
   A \hspace{70 mm} B
\caption{A) Phase diagram of the heavy-quark model in the \((\mu,T)\)-plane. The isotropic case is shown with \(c_B = 0\) and magnetized anisotropic cases are denoted with different \( c_B \). The magenta stars indicate the CEPs.  B) The critical transition temperature at zero chemical potential, $\mu=0$, as a function of the magnetic field parameter $c_B$. We set $q_B=5$. 
}
 \label{Fig:Tmuhq}
\end{figure}

The phase diagram (the 1st order phase transition location) for the heavy-quark model in the \((\mu,T)\)-plane, considering the isotropic case with \(c_B = 0\) and magnetized anisotropic cases with different \( c_B \) is shown in Fig.\,\ref{Fig:Tmuhq}A. In this figure, the magenta stars indicate the corresponding critical end points (CEPs) for $c_B=0$ and $c_B=-0.5$ GeV${}^2$. 
Increasing the absolute value of  the magnetic field parameter $c_B$,  the length of the phase transition curves in the \((\mu,T)\)-plane increases up to $c_B=-0.5$ GeV${}^2$ and then decreases. The critical transition temperature $T_c$ (the 1st order phase transition temperature) at zero chemical potential, $\mu=0$, as a function of the magnetic field parameter $c_B$ is depicted in Fig.\,\ref{Fig:Tmuhq}B. This figure shows that $T_c$ increases with increasing absolute value of the magnetic field parameter $c_B$ and confirms the magnetic catalysis phenomenon for the heavy-quark model.

\begin{figure}[h!]
\centering
\includegraphics[scale=0.475]{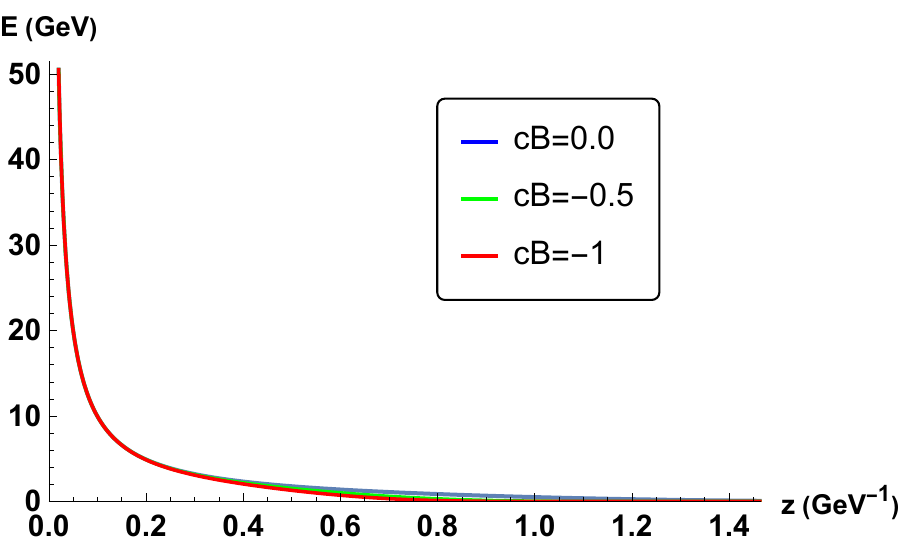} 
\includegraphics[scale=0.475]{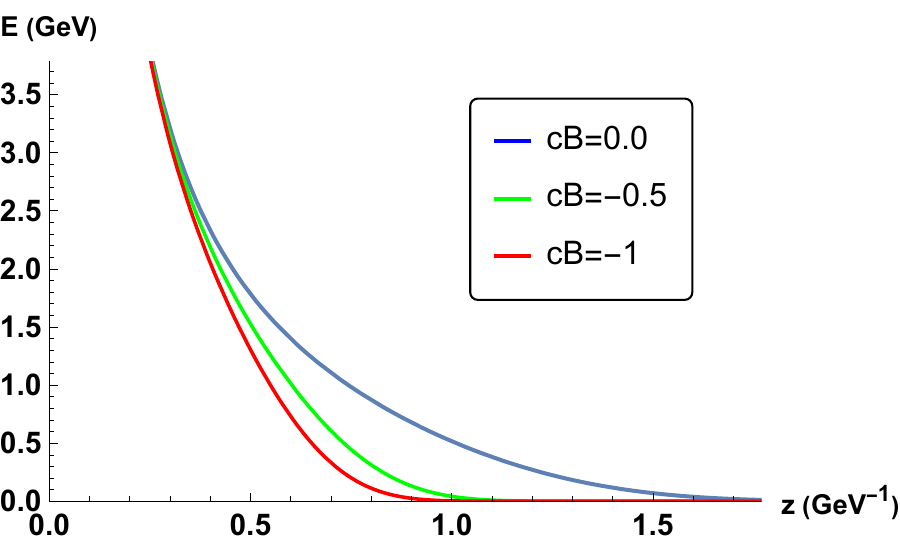} \\
   A \hspace{60 mm} B
\caption{A) Energy scale \( E \) (GeV) in the boundary field theory of the heavy-quark model as a function of the holographic coordinate \( z \) (GeV${}^{-1}$), corresponding to the warp factor \( \fb(z) \) for different values of $c_B$. B) Zoom in of the left panel. We set $q_B=5$; \( [c_B] = \) GeV${}^2$.
}
\label{Fig:E-z-HQ}
\end{figure}

The energy scale \( E \) of the boundary field theory corresponds to the prefactor of the metric  \eqref{eq:2.03} \cite{Galow:2009kw} (for more details see refs. in \cite{Arefeva:2024xmg}). In other words, the energy scale \( E \) (GeV) as a function of the holographic coordinate \( z \) (GeV${}^{-1}$) for heavy-quark model is given by  
\be \label{BBL}
E = \frac{\sqrt{\fb(z)}}{z} =\frac{e^{-\fc\,z^2/4-(p-c_B\, q_B) z^4}}{z}\, ,
\ee
where the parameters $\fc$ and $p$ have already been introduced in \eqref{scaleHQX}. The energy scale $E(z)$ is shown in Fig.\,\ref{Fig:E-z-HQ}A and a zoom in of the left panel is shown in Fig.\,\ref{Fig:E-z-HQ}B for different values of the magnetic field parameter $c_B$. 

Fig.\,\ref{Fig:E-z-HQ}B illustrates that the energy scale 
$E$ in the boundary field theory depends not only on the holographic coordinate 
$z$ but also on the magnetic field parameter 
$c_B$. Specifically, as the absolute value of 
$c_B$
 increases, the energy scale 
$E$
 decreases more rapidly with respect to 
$z$. For a fixed value of 
$z$ on the gravity side, a larger 
$c_B$ corresponds to a lower energy scale 
$E$ on the gauge theory side.

\begin{figure}[h!]
  \centering
   \includegraphics[scale=0.37]{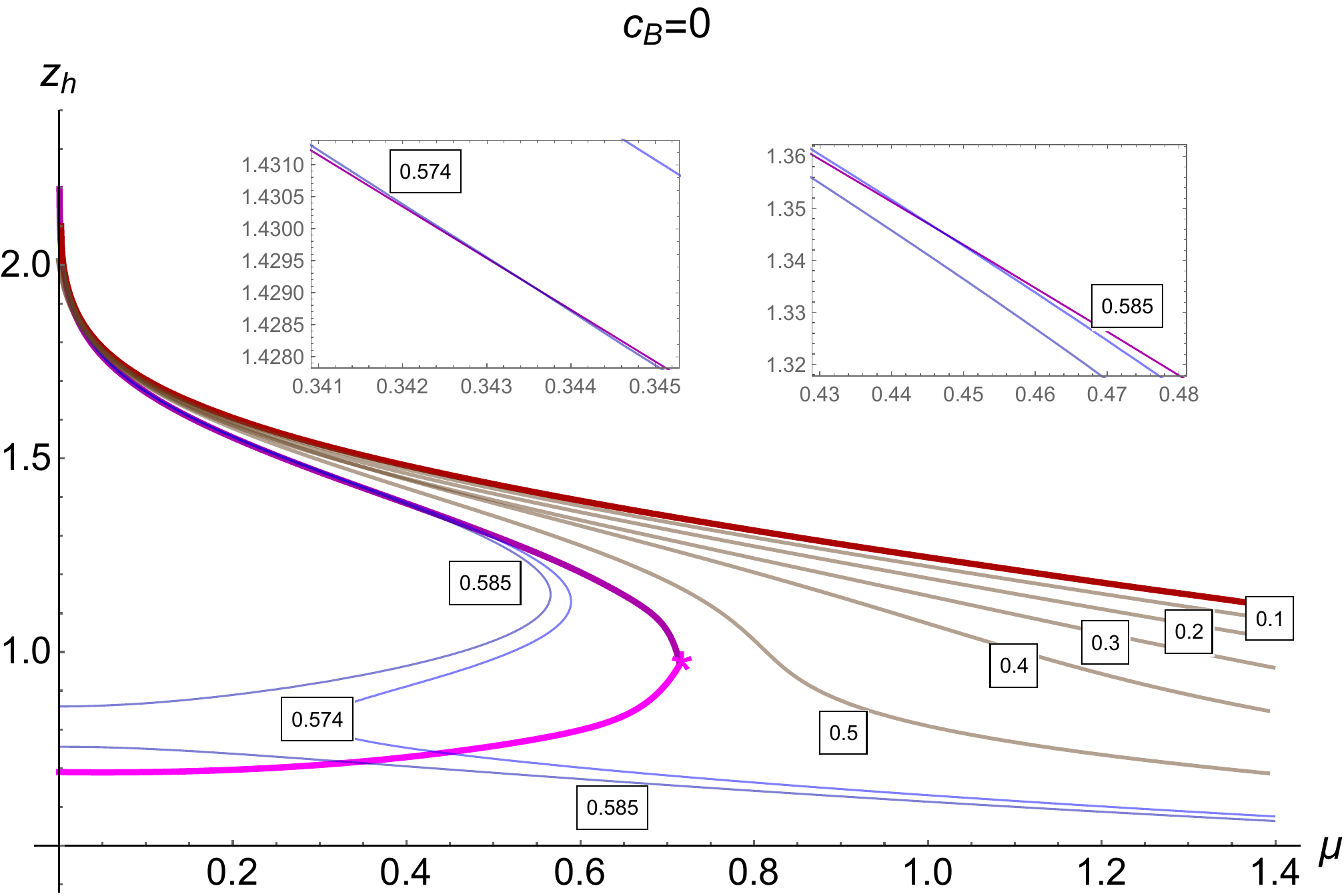}\\
   \vspace*{3.4mm}
  \includegraphics[scale=0.3]{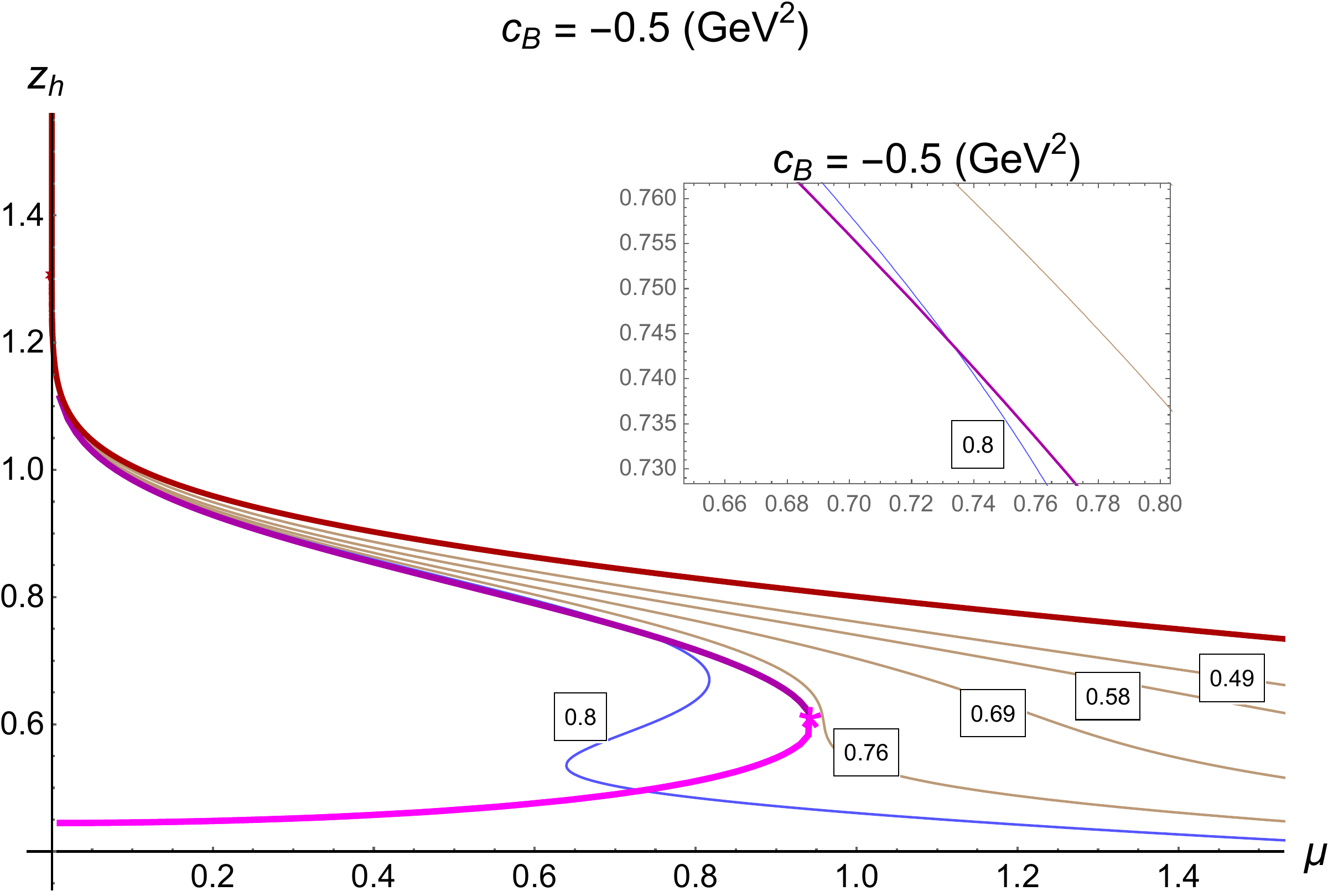} 
\caption{2D plots in the \((\mu,z_h)\)-plane
 for $c_B = 0$
(top panel) and $c_B = -0.5$ GeV${}^2$
(bottom panel). 1st order phase transition lines are shown as magenta curves, with CEPs marked by magenta stars. The domain between magenta lines indicates the unstable domains. Fixed-temperature contours are displayed in brown when they do not intersect the 1st order phase transition lines and in blue where they cross the 1st order transitions. Corresponding temperatures 
$T$
(in GeV) are annotated in white boxes. The 
$T=0$
contours are highlighted in dark red. Zoomed regions illustrate intersections between the blue contours and the 1st order transition line. We set $q_B=5$; $[\mu]=[z_h]^{-1} =$ GeV.
}
  \label{Fig:Z_h-mu-approxHQ}
\end{figure}

 2D plots in the 
$(\mu,z_h)$-plane are shown for 
$c_B=0$ in Fig.\,\ref{Fig:Z_h-mu-approxHQ} (top panel), while Fig.\,\ref{Fig:Z_h-mu-approxHQ} (bottom panel) displays the corresponding plots for 
$c_B=-0.5$ GeV$^2$. The magenta curves represent 1st order phase transitions, consisting of dark and light branches. Fixed temperatures are indicated by brown and blue contours, with 
$T=0$ represented by dark red contours. The dark and light branches of the magenta curve are connected at the critical endpoint (CEP), whose coordinates ($\mu_c,T_c$)
 depend on the magnetic field parameter 
$c_B$. Based on Fig.\,\ref{Fig:Tmuhq}, the CEP in Fig.\,\ref{Fig:Z_h-mu-approxHQ}B is located at ($\mu_c,T_c$)=($0.94,0.76$).

The region above the dark branch of the magenta curve and below the light branch corresponds to domains of stable black hole solutions, which represent physical domains. In contrast, the region between the light and dark magenta branches corresponds to a nonphysical domain or unstable solutions. The area between the dark magenta line and the dark red line predominantly represents the hadronic phase, while the region just below the light magenta line corresponds to the quarkyonic phase. For low $z_h$, the quarkyonic phase transitions into the quark-gluon plasma (QGP) phase (see Fig. 15 of \cite{Arefeva:2024vom} for a more detailed illustration).

\newpage
$$\,$$

\section{Running Coupling in a Magnetized Background }\label{sec:running}

In the holographic approach, the running coupling \( \alpha \) is a function of holographic coordinate $z$, i.e. \( \alpha(z) \) is defined in terms of the dilaton field \( \varphi(z) \)  \cite{Gursoy:2007cb,Gursoy:2007er,Pirner:2009gr}.
\be\label{lambda-phi}
\alpha(z) = e^{\varphi(z)} \,.
\ee

To obtain the EOM of the dilaton field, we need to solve the system of EOMs \eqref{eq:2.05}-\eqref{eq:2.10}. Solving the EOMs needed to fix the form of the scale factor \eqref{scaleHQX} and gauge coupling function $f_1$ \cite{Arefeva:2023jjh}. We can consider the following  gauge coupling function $f_1$ for the ``heavy-quark'' model including magnetic catalysis

\begin{gather}
  f_1 = e^{-(\frac{2}{3}R_{gg}+\frac{c_B q_B}{2})z^2+(p-c_{B}q_{B})z^4} \,.
  \label{eq:2.14}
\end{gather}
To obtain the solution for the dilaton field we need to solve EOM  \eqref{eq:2.08} with the scale factor \eqref{scaleHQX}. We obtain the following expression 
\begin{gather}
  \begin{split}
    \varphi (z) &= \int_{z_0}^z\, \frac{d\xi}{ \xi} \Biggl[
    - \, 4 + \frac{2}{3} \Biggl(
    6 + 3 \, (- c_B + 6 R_{gg}) \, \xi^2
    + \bigl(- \, 3 \, c_B(c_B + 60 \, q_B) \, + \\
    &+ 4 (45 \, p + R_{gg}^2) \bigr) \, \xi^4
    + 48 R_{gg} (p - c_B \, q_B) \, \xi^6 
    + 144 \, (p-c_B \, q_B)^2 \, \xi^8 \Biggr)
    \Biggr]^{\frac{1}{2}} , \label{eq:4.80}
 \end{split}
  \end{gather}
where $z_0$ is fixed from the boundary condition for the dilaton field  defined below.

Note that the dilaton field has a crucial role in defining the holographic running coupling. In addition, the dilaton field \( \varphi(z) \) has an integration constant and needs to apply a boundary condition to be uniquely determined. Therefore, utilizing different boundary conditions lead to different physical results. 
The boundary condition can be chosen in the following form:
\be \label{zerobc2}
\varphi_{z_0}(z)\Big|_{z=z_0} = 0\,.
\ee
where $z_0$ is an arbitrary holographic coordinate.
The particular form of the boundary condition that corresponds to $z_0=0$ is 
\be \label{zerobc1}
\varphi_0(z)\Big|_{z=0} = 0\,.
\ee
 Considering the modified boundary condition \( z_0 = \fz(z_h) \), the running coupling is given by:
\be
\alpha_{\fz}(z; T, \mu) =  e^{\varphi_0(z)-\varphi_0(\fz(z_h))}.
\ee
See more details about  the running coupling definition in \cite{Arefeva:2024xmg,Arefeva:2024vom}.  The physical boundary condition for the dilaton field that we consider to study the heavy-quark model \cite{Arefeva:2024vom} is given by
\begin{equation}\label{LQz0}
z_0 = \fz(z_h) = \exp{(-\frac{z_h}{4})} + 0.1 \,, 
\end{equation}
where the dilaton field becomes zero at \( z=z_0 \), as specified in \eqref{zerobc2}. Note that the physical boundary condition in \eqref{LQz0}, that was utilized to study the beta-function and renormalization group flow in  \cite{Arefeva:2025xtz}, was derived by investigating the behavior of the QCD string tension as a function of the temperature at zero chemical potential for the isotropic case. For more details see \cite{Arefeva:2024vom}.
In spite of the fact that lattice results are not available for the anisotropic case, i.e. \( c_B \neq 0 \), the physical boundary condition \eqref{LQz0} is still used in this research.

It is important to  note that in studying the behavior of the running coupling constant at different energy scales $E$, and different magnetic field parameters $c_B$, we need to respect the physical domains of the model in Fig.\,\ref{Fig:Z_h-mu-approxHQ}. The physical domains in \((\mu,z_h)\)-plane for nonzero chemical potential are determined by the sizes of the black hole horizons $z_h$, corresponding to the 1st order phase transition between small and large black holes, denoted by the magenta lines.

\begin{figure}
  \hspace*{-0.7cm}
  \includegraphics[scale=0.18]{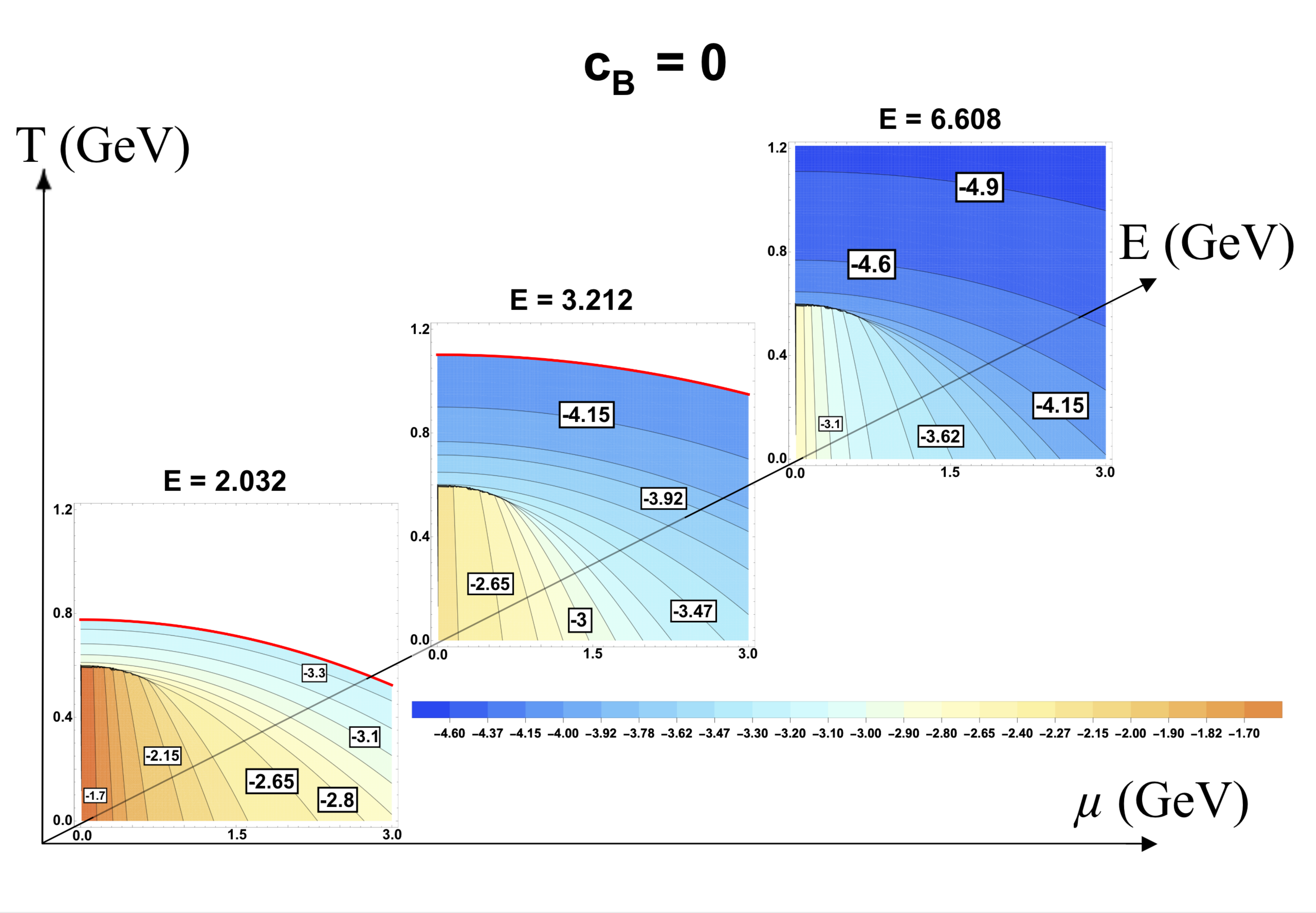}
\caption{Density plots with contours for $\log\alpha_{\fz}(E; \mu, T)$ at different energy scales  $E = \{2.032, 3.212, 6.608 \} $ (GeV) for $c_B =0$. All values of $E$ on the top of each panel show the fixed value of the energy $E$-coordinate. The red lines in the first two graphs on the left indicate the maximum possible temperature values at the corresponding energy scale $E$.
}
\label{Fig:phi-cb0}
\end{figure}

Fig.\,\ref{Fig:phi-cb0} shows density plots with contours for the logarithm of the running coupling \( \log\alpha_{\fz}(E; \mu, T) \) for the heavy quark model at \( c_B = 0 \) considering different energy scales \( E = \{2.032, 3.212, 6.608 \} \) (GeV) in the boundary field theory. Each panel is depicted at a fixed value of energy  $E$-coordinate, which is represented on top of them. The maximum possible temperature values at the corresponding energy scale \( E \) are denoted by the red lines in the first two graphs, which arise to fulfill the physical condition, i.e.  \( z \leq z_h \). The fixed values of energy scales in Fig.\,\ref{Fig:phi-cb0} correspond to holographic coordinates \( z = \{0.436, 0.295, 0.150\} \) (GeV${}^{-1}$) represented in Fig.\,\ref{Fig:E-z-HQ}. Different but fixed values of \( \log\alpha_{\fz} \)
corresponding to different contours in each panel are shown with associated values in the rectangles. From  Fig.\,\ref{Fig:phi-cb0} it is clear that as the energy scale increases, the contours corresponding to the fixed values of the running coupling shift to lower values of the temperature. Therefore, increasing the temperature, the running coupling decreases monotonically, indicating that for a fixed given temperature, the running coupling decreases as the energy scale increases.

\begin{figure}
  \hspace*{-0.7cm}
  \includegraphics[scale=0.18]{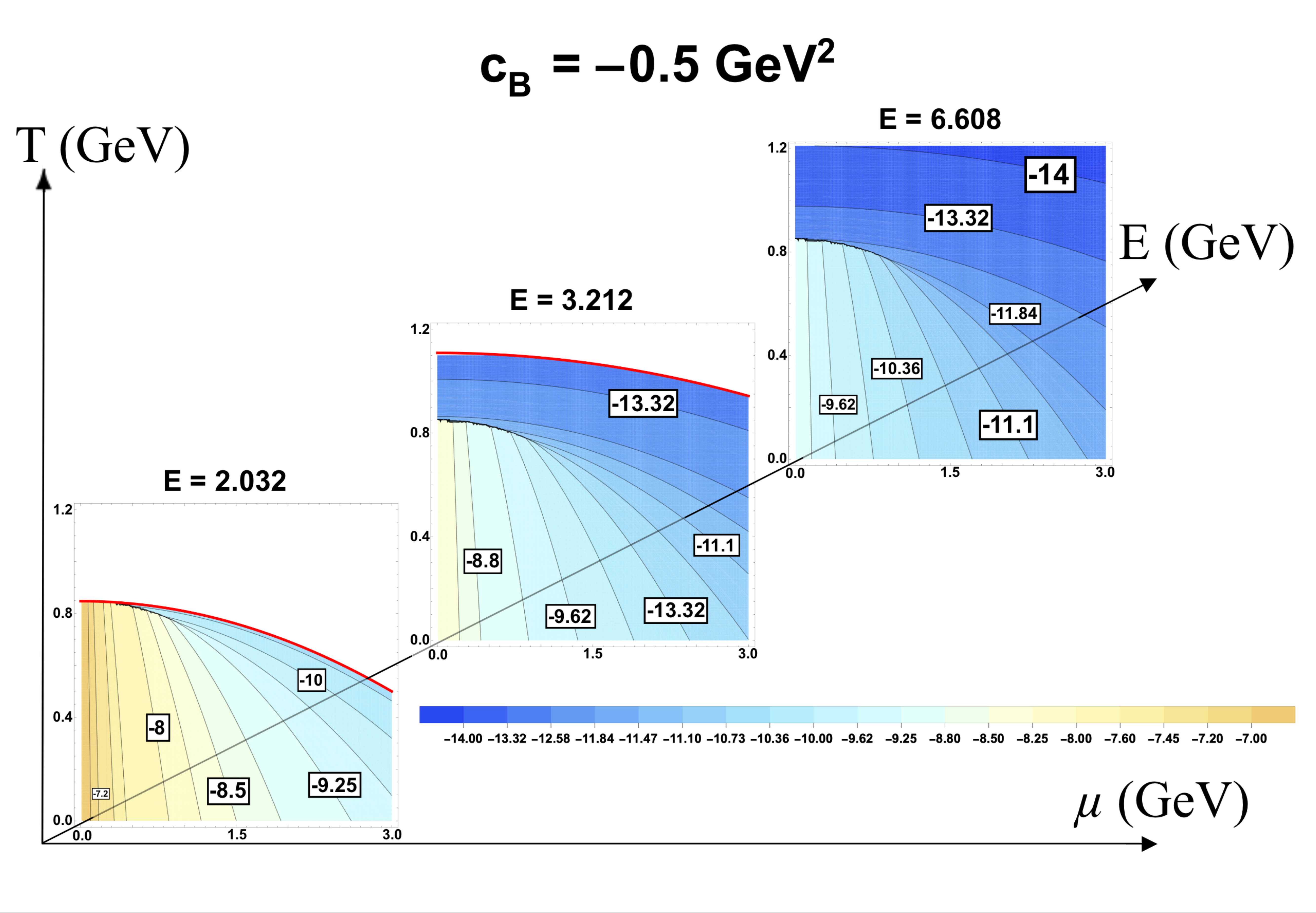}
\caption{Density plots with contours for $\log\alpha_{\fz}(E; \mu, T)$ at different energy scales $ E = \{2.032, 3.212, 6.608 \}$ (GeV) are shown for $ c_B=-0.5$ GeV${}^2$. All values of $E$ on the top of each panel show the fixed value of the energy $E$-coordinate. These plots illustrate the variation of the running coupling constant across different energy scales, chemical potentials, and temperatures in the presence of a magnetic field.  We set $q_B=5$.}
\label{Fig:phi-cb05}
\end{figure}

In Fig.\,\ref{Fig:phi-cb05}, density plots with contours for \( \log\alpha_{\fz}(E; \mu, T) \) are depicted line in Fig.\,\ref{Fig:phi-cb0}, but considering \( c_B = -0.5 \) GeV${}^2$. 
The fixed values of energy scales in Fig.\,\ref{Fig:phi-cb05} correspond to holographic coordinates \( z = \{0.421, 0.294, 0.150\} \) (GeV${}^{-1}$) represented in Fig.\,\ref{Fig:E-z-HQ}. Here, the $z$ values are different from $c_B=0$ because the energy scale $E$ for some domain of energy depends on the magnetic field parameter $c_B$ as well. 

The running coupling constant $\alpha$, as shown in Figs.\,\ref{Fig:phi-cb0} and \ref{Fig:phi-cb05} shows minimal dependence on the temperature in the hadronic phase for both isotropic ($c_B=0$)
 and anisotropic ($c_B=-0.5$ GeV${}^2$) cases. However, the running coupling $\alpha$ shows significant dependence on the temperature and chemical potential in the QGP phase for both isotropic ($c_B=0$)
 and anisotropic ($c_B=-0.5$ GeV${}^2$) cases.

The comparison of Fig.\,\ref{Fig:phi-cb0} ($c_B=0$) and Fig.\,\ref{Fig:phi-cb05} ($c_B=-0.5$ GeV${}^2$) reveals that  at fixed values of the parameters of the model ($E; \mu; T$) the strength of the running coupling decreases in the presence of the magnetic field.

\newpage
$$\,$$
\newpage

\section{Conclusion} \label{Sect:Conc}

In this paper we investigate the behavior of the running coupling constant in an anisotropic holographic model under a strong magnetic field for the heavy-quark case \cite{Arefeva:2024vom}.  Note that, in \cite{Arefeva:2024xmg} the influence of magnetic field on the light quark running coupling was studied. Both models are constructed using the Einstein-dilaton-two-Maxwell action and incorporate a 5-dimensional metric with a warp factor, extending earlier isotropic frameworks—specifically, the heavy-quark model from \cite{Andreev:2006ct,He:2010ye,Arefeva:2018hyo,Bohra:2019ebj,Arefeva:2020vae,Yang:2015aia,Bohra:2020qom,Arefeva:2023jjh} and the light-quark model in \cite{Li:2017tdz,Chen:2020ath,Arefeva:2022avn}.
\\

The model of heavy quarks in an external magnetic field has both similarities and significant differences
with the model of light quarks
\begin{itemize}
\item A key distinction lies in the response to the magnetic field: the heavy-quark model exhibits magnetic catalysis, whereas the light-quark model demonstrates inverse magnetic catalysis. This contrast highlights the interplay between  magnetic field strength and quark mass in shaping the phase structure of these holographic  models.
\item The distinct positions of the 1st order phase transition lines in the two models lead to contrasting behaviors in the discontinuity of 
$\alpha$
 across the 1st order phase transition:
 \begin{itemize}
\item For the light-quark model, the jump in 
$\alpha$
 increases with increasing  $\mu$.

\item For the heavy-quark model, the jump decreases with increasing $\mu$.
\end{itemize}
\end{itemize}

This implies that the CEP is approached by moving along the phase transition line  toward higher $\mu$
 in the heavy-quark model, but toward lower 
 $\mu$
in the light-quark model.

For both isotropic ($c_B=0$)
 and anisotropic ($c_B=-0.5$ GeV${}^2$):
\begin{itemize}
\item For light-quark model, in the hadronic phase,  $\alpha$ shows minimal variation with changes in $\mu$.
\item For heavy-quark model, in the hadronic phase,  $\alpha$ shows minimal variation with changes in $T$.
\end{itemize}

But there are also some similarities. Namely, the heavy-quark model under an external magnetic field shares the following qualitative similarities with the light-quark model:
\begin{itemize}
\item At fixed temperature 
$T$
 and chemical potential 
$\mu$, the magnetic field suppresses the running coupling constant 
$\alpha$.
\item The running coupling 
$\alpha$
decreases monotonically with increasing energy scale 
$E$ (at fixed 
$T$
 and 
$\mu$), consistent with asymptotic freedom. This trend remains robust even in magnetized systems.

\item In the QGP phase  for light or heavy quarks model, 
$\alpha$ demonstrates pronounced sensitivity to 
$T$
 and 
$\mu$, regardless of anisotropy ($c_B=0$ or
$c_B\neq 0$).\\
\end{itemize}

It is important to note that heavy and light quarks model exhibit very different characteristics in the presence of the magnetic field \cite{Arefeva:2024xmg,Arefeva:2022avn,Arefeva:2023jjh}. The holographic  QCD running coupling does depend on the mass of the quark. 
In holographic approach running coupling is defined in terms of dilaton field, see Eq.\,\eqref{lambda-phi} and the dilaton field depends of the choice of warp factor, see Eq.\,\eqref{eq:2.08} that is different for heavy and light quarks. Therefore, running coupling is considered for heavy and light quarks separately.
This point has been investigated in the Fig.\,34 \cite{Arefeva:2024vom} for the light quark running coupling and in the Fig.\,44 \cite{Arefeva:2024vom} for the heavy quark running coupling via bottom-up holographic approach.\\

 Comparisons between holographic results and 2-loop pQCD calculations are provided in \cite{He:2010ye,He:2010bx,Galow:2009kw}. For a detailed analysis of the correspondence between the pQCD and holographic beta functions under varying $n_f$, see \cite{Arefeva:2024poq}.\\

A promising direction for future research would be to systematically investigate how magnetic fields modify the holographic beta-function in holographic QCD frameworks, particularly in comparison to the zero-field case considered in \cite{Arefeva:2025xtz}.
In addition, considering heavy and light quark models as a unified model, i.e. hybrid model is one of our future goals to achieve.\\

\section*{Acknowledgments}

We thank K. Rannu and M. Usova for their valuable discussions. 
The work of I. A., P. S. and A.N. is supported by Theoretical Physics and Mathematics Advancement Foundation ``BASIS” (grant No. 24-1-1-82-1, grant No. 23-1-4-43-1 and grant No. 24-2-2-4-1, respectively).
The work of A. H. was performed at the Steklov International Mathematical Center and supported by the Ministry of Science and Higher Education of the Russian Federation (Agreement No. 075-15-2022-265).

\newpage
\appendix
{\large \bf Appendix}

\section{Solution to EOMs }\label{app1}

 By varying the action \eqref{eq:2.01}, the following EOMs can be  obtained

\begin{gather}
 \begin{split}
    \varphi'' &+ \varphi' \left( \cfrac{g'}{g} + \cfrac{3 \fb'}{2 \fb} -
      \cfrac{3}{z} + c_B z \right)
    + \left( \cfrac{z}{L} \right)^2 \cfrac{\partial f_1}{\partial
      \varphi} \ \cfrac{(A_t')^2}{2 \fb g} - \left( \cfrac{z}{L} \right)^{2} \cfrac{\partial
      f_B}{\partial \varphi} \ \cfrac{q_B^2}{2 \fb g} \\ &
    \qquad \qquad \qquad \qquad \qquad - \left( \cfrac{L}{z} \right)^2 \cfrac{\fb}{g} \ \cfrac{\partial
      V}{\partial \varphi} = 0\,,
  \end{split}
  \label{eq:2.05} \\
      A_t'' + A_t' \left( \cfrac{\fb'}{2 \fb} + \cfrac{f_1'}{f_1} -
    \cfrac{1}{ z} + c_B z \right) = 0\,, \label{eq:2.06} \\
    g'' + g' \left(\cfrac{3 \fb'}{2 \fb} - \cfrac{3}{ z} + c_B
    z \right)
  - \left( \cfrac{z}{L} \right)^2 \cfrac{f_1 (A_t')^2}{\fb}
  - \left( \cfrac{z}{L} \right)^{2} \cfrac{q_B^2 \
    f_B}{\fb} = 0\,, \label{eq:2.07}
\\  
  \fb'' - \cfrac{3 (\fb')^2}{2 \fb} + \cfrac{2 \fb'}{z}
 + \cfrac{2 \fb\, c_B}{3 } \left( 1
    + c_B z^2 \right)
  + \cfrac{\fb \, (\varphi')^2}{3} = 0\,, \label{eq:2.08} 
  \\
  \begin{split}
   c_B z^2  \left(2 g'+3 g \right) 
      \left(
        \cfrac{\fb'}{\fb} - \cfrac{4}{3 z} + \cfrac{2 c_B z}{3}
      \right) 
    - \left( \cfrac{z}{L} \right)^{3}
    \cfrac{L \,q_B^2 f_B }{\fb} = 0\,, 
  \end{split}\label{eq:new} \\
  \begin{split}
    \cfrac{\fb''}{\fb} &+ \cfrac{(\fb')^2}{2 \fb^2}
    + \cfrac{3 \fb'}{\fb} \left( \cfrac{g'}{2 g}
      - \cfrac{2}{ z}
      + \cfrac{2 c_B z}{3} \right)
    - \cfrac{g'}{3 z g}  \left(9 - 3 c_B z^2
    \right)  - \cfrac{2 c_B}{3} \left(  5-c_B z^2 \right)\\
    & \qquad \qquad \qquad \qquad \quad + \cfrac{8}{ z^2} 
    + \cfrac{g''}{3 g} + \cfrac{2}{3} \left( \cfrac{L}{z} \right)^2
    \cfrac{\fb V}{g} = 0\,,
  \end{split}\label{eq:2.10}
\end{gather}
where, the symbol "$'$" denotes differentiation with respect to the holographic coordinate $z$. The EOMs \eqref{eq:2.05}–\eqref{eq:2.10} have a general form and can be applied to both heavy and light-quark models. Although the EOMs contain six equations, \eqref{eq:2.05} is the consequence of other equations and then we have just five independent EOMs. Note  that all functions depend on the holographic coordinate $z$.

To solve the EOMs, we apply the following boundary conditions:
\begin{gather}
  A_t(0) = \mu, \quad A_t(z_h) = 0, \label{eq:2.11} \\
  g(0) = 1, \quad g(z_h) = 0, \label{eq:2.12} \\
  \varphi(z_0) = 0\, , \label{eq:2.13}
\end{gather}
where $z_0$ and $z_h$ were introduced and defined in section  \eqref{sec:prelim}.

We emphasize that to obtain the exact solutions for EOMs \eqref{eq:2.05}–\eqref{eq:2.10} also we need to use the proper ansatz for the gauge coupling function $f_1$ that has been already introduced in \eqref{eq:2.14}. In addition, it is needed to fix the functionality of  the scale factor $\cA(z)$ for heavy quarks which was introduced in \eqref{scaleHQX}. 
\\

\newpage

\end{document}